\documentstyle[12pt]{article}

\def\ref#1{$^{#1)}$}
\begin{document}
\begin{titlepage}
\begin{center}
\today     \hfill    LBL-33677 \\
           \hfill    UCB-PTH-93/04 \\

\vskip .25in

{\large \bf Towards A Theory Of Quark and Lepton Masses.}
\footnote{ Plenary talk given at the 16th Texas Symposium on Relativistic
Astrophysics and 3rd Symposium on Particles, Strings and Cosmology, Berkeley,
December 1992.}
\footnote{ This work was supported in part by the Director, Office of
Energy Research, Office of High Energy and Nuclear Physics, Division of
High Energy Physics of the U.S. Department of Energy under Contract
DE-AC03-76SF00098 and in part by the National Science Foundation under
grant PHY90-21139.}

\vskip .25in

L. J. Hall\\[.25in]

{\em  Department of Physics, University of California\\
      and\\
      Theoretical Physics Group, Physics Division\\
      Lawrence Berkeley Laboratory\\
      1 Cyclotron Road, Berkeley, California 94720}
\end{center}

\vskip .25in

\begin{abstract}
Has any progress been made on understanding and predicting
the 13 parameters which describe the observed masses and mixing
angles of the quarks and leptons?
Arguments are given in favor of pursuing schemes in which grand unified and
family symmetries provide many relations among these 13 parameters.
A sequence of simple assumptions leads to a supersymmetric SO(10) theory with
8 predictions: $\tan \beta, m_t, V_{cb}, m_s, m_s/m_d$, $m_u/m_d, V_{ub}$ and
the amount of CP violation $J$. These predictions are presented, together
with experiments which will test them.
\end{abstract}
\end{titlepage}
\renewcommand{\thepage}{\roman{page}}
\setcounter{page}{2}
\mbox{ }

\vskip 1in

\begin{center}
{\bf Disclaimer}
\end{center}

\vskip .2in

\begin{scriptsize}
\begin{quotation}
This document was prepared as an account of work sponsored by the United
States Government.  Neither the United States Government nor any agency
thereof, nor The Regents of the University of California, nor any of their
employees, makes any warranty, express or implied, or assumes any legal
liability or responsibility for the accuracy, completeness, or usefulness
of any information, apparatus, product, or process disclosed, or represents
that its use would not infringe privately owned rights.  Reference herein
to any specific commercial products process, or service by its trade name,
trademark, manufacturer, or otherwise, does not necessarily constitute or
imply its endorsement, recommendation, or favoring by the United States
Government or any agency thereof, or The Regents of the University of
California.  The views and opinions of authors expressed herein do not
necessarily state or reflect those of the United States Government or any
agency thereof of The Regents of the University of California and shall
not be used for advertising or product endorsement purposes.
\end{quotation}
\end{scriptsize}

\vskip 2in

\begin{center}
\begin{small}
{\it Lawrence Berkeley Laboratory is an equal opportunity employer.}
\end{small}
\end{center}

\newpage
\renewcommand{\thepage}{\arabic{page}}
\setcounter{page}{1}

A major component of experimental high energy physics is the
measurement of the masses of the quarks and leptons and their
couplings to the $W$ boson.
There is no mystery about why this is so: we are interested in learning the
fundamental parameters of the standard model, and 13 out of 18 of these
correspond to quark and lepton masses and mixing.
I am not trying to minimize the importance of the 5 parameters of the gauge
and Higgs
sectors, which can be taken as $\alpha, \alpha_s, M_Z, G_F$ and $M_H$; but it
is
a  simple fact that the majority of the fundamental parameters belong to
the flavor sector. These 13 parameters consist of 9 masses: for the up-type
quarks $m_u, m_c$ and $m_t$, the down-type quarks $m_d, m_s$ and $m_b$, and
the charged leptons $m_e, m_\mu$ and $m_\tau$; and the 4 independent parameters
of the Kobayashi-Maskawa (KM) mixing matrix, which I take as the Cabbibo angle
$\sin \theta_c = |V_{us}|$,  $|V_{cb}|$,  $ |V_{ub}|$ and the parameter $J$
which describes the amount of CP violation in the KM matrix.

Each of the 18 fundamental parameters is represented in the standard model by a
coupling constant.
I do not know how to construct a fundamental theory and perform a first
principles calculation of these coupling constants.
Does this mean I have no hope of making predictions?
{\bf No}.
It is always possible to obtain predictions by {\it reducing the number of free
parameters}.
The Balmer formula provides an illustration of this.
A large number of observables (the hydrogenic spectral wavelengths) are
described by a single free parameter (the Rydberg constant).
Twenty eight years after this incredibly successful formula was written down,
it
was understood by Bohr; indeed, his atomic model gave a theoretical prediction
for the Rydberg constant, $ R = 2\pi^2 mZ^2e^4/h^3$.
This crowning achievement was the birth of the quantum theory of atomic
structure.
It may well be that a predictive scheme for fermion masses, depending on far
fewer than the 13 flavor couplings of the standard model, is a prerequisite for
the development of a fundamental theory of fermion masses.

It is interesting to recall in a little more detail how the development of
atomic theory and quantum mechanics grew out of studies of spectral
wavelengths, and to compare this evolution with the spectroscopy of today: that
of quark and lepton masses. I would argue that the development proceeds from
the
experimental measurements of the phenomena, first to a recognition of
regularities amongst the measurements, then to the physical insight which gives
some understanding of these regularities and finally to a fundamental theory,
which allows the totality of the phenomena to be understood from a few general
principles. The spectral lines of hydrogen were first measured accurately, to 1
part in $10^4$ for four of the lines of the Balmer series, by Angstrom
in  1860.
A simple regularity in this series of wavelengths was noticed by Balmer in
1885, which he described by the formula $\lambda = C { n^2 \over n^2 + 4}$
where n is an integer and C a constant. Few
would disagree that the great leap of physical insight came from Bohr, with his
atomic model of 1913. This provided a picture of what was going on
in terms of discrete energy levels, together
with a derivation of Balmer's formula in the now more familiar form
$\nu = R \left( {1 \over n^2_1} - { 1 \over n_2^2} \right) $ and a theoretical
prediction for the one parameter $R$ in terms of $\alpha$ and $m_e$. Over the
next fifteen years this led to the development of quantum mechanics, a radical
new foundation underlying all physics.

At what stage of the development process do we stand today with regard to quark
and lepton spectroscopy? I would guess that if you asked this to a
cross-section of particle physicists, most would say that we are somewhere
between Angstrom (1860) and Balmer (1885). We have some reasonable data, but
essentially no understanding of the regularities or of the underlying theory.
This may indeed be the situation. Attempts to predict the quark and lepton
masses in gauge theories began in 1972 \cite{WGG},
immediately after these theories were shown to be
renormalizable (ie predictive), and there have been a variety of approachs,
each with an interesting history. Some
schemes have been very ambitious, suggesting an origin for fermion masses very
different than the description provided by the standard model. Two such
examples are extended technicolor \cite{etc} and string theory \cite{string};
however, despite
considerable effort, it is not known whether these ideas are consistent with
the
observed masses, and they are certainly
very far from providing predictive relations that
can be tested. The same criticism cannot quite be leveled at the scheme known
as top condensates \cite{tcond} as this does predict the top quark and Higgs
boson masses. Nevertheless, while it would be interesting to find
$m_t$ near 230 GeV and $m_H$ near 260 GeV, it would hardly be persuasive,
because these results correspond to infrared fixed points and are therefore
quite insensitive to the underlying physics at high mass scales.
There are other ideas, such as gauged generation symmetries and radiative
hierarchies, which are very well motivated, but which again have not led to
concrete accurate predictions which can be experimentally tested. One can
therefore argue quite persuasively that, not only are we far from having a
theory of fermion masses, but many avenues are open precisely because the
regularities of the quark and lepton masses have yet to be found.

In this talk I would like to argue the case for an alternative viewpoint: that
we can already see and understand some of the regularities. This viewpoint may
be completely mistaken, but it should be taken seriously because it is the only
direction which has provided a sufficient number of accurate predictions to
qualify as being ``testable''. This direction is the one of {\em parameter
reduction obtained by imposing symmetries.} If this viewpoint is correct, our
stage of development is somewhere between Balmer (1885) and Bohr (1913),
perhaps even close to Bohr.

The regularities are not embodied in a simple formula of the Balmer type, but
in
the framework of todays tools of theoretical physics: symmetries. Infact, such
successful predictions have only been obtained by the combination of three very
different types of symmetry. The first, grand unified gauge symmetry is both
very elegant and very powerful. It allows relations between the up, down and
lepton masses. The second, family symmetry, is also very powerful leading to a
substantial parameter reduction, however at the moment it is very ad hoc and is
the weak link in the chain. The final symmetry, supersymmetry, actually leads
to an increase in the number of parameters, but is apparently required by data
since otherwise many of the predictions are not correct. Although there is a
simple group theoretic understanding of the Yukawa coupling structure, these
simple regularities are not immediately manifest in the observed masses
because the Yukawa couplings are modified by calculable dynamical effects.

There is an experimental hint that the above viewpoint of parameter reduction
in the flavor sector is worth pursuing: progress has been made
in reducing the number of parameters in the gauge sector.
In grand unified theories (GUTs) the three independent gauge couplings become
related \cite{GG}.
This implies predictions for the weak scale gauge couplings $g_i(M_W)$,
i = 1,2,3 of the form \cite{GQW}:
$$
g_i(M_W) = C_i \; g_G  \; \eta_i \eqno(1)
$$
where $g_G$ is the GUT gauge coupling, $C_i$ are numerical
group theory constants and the $\eta_i$, which are radiative corrections
computed with the renormalization group, depend on mass ratios such as
$M_W/M_G$, where $M_G$ is the GUT scale.
How many predictions occur in the gauge sector of GUTs?
While the $C_i$ are purely numerical group theory constants,
the $\eta_i$ depend on ratios
of various mass scales.
If there are two or more mass ratios on which the $\eta_i$ depend, then there
are no predictions: together with $g_G$ there are three or more free parameters
for the three standard model parameters $g_i$.
The only hope is for the maximally predictive possibility that the $\eta_i$
depend only on the single mass ratio $M_W/M_G$, in which case there will be one
prediction, usually chosen to be the weak mixing angle $\sin^2\theta$.

There are many possible GUTs which have no new scale other than $M_G$.
How many different predictions for $\sin^2\theta$ can they give?
The answer is basically just two: .211 without supersymmetry and .233 with
weak-scale supersymmetry \cite{susys2}.
What is the accuracy of these predictions?
There are threshold corrections from GUT \cite{gutth}, Planck \cite{planckth},
and weak scales \cite{weakth}
which are typically around $.002$.
Since the standard model is consistent with any value of $\sin^2\theta$
from 0 to 1, I think that it is very significant that the minimal
supersymmetric
scheme predicts precisely the experimental value of .233 $\pm$ .001.
Many people shrug this off, pointing out that it is just one number. However
{\em it
is the only significant prediction} of any of the 18 parameters of the standard
model, and hence I take it as a valuable indication from experiment that these
theories are worth pursuing further. In particular it lends
support for two of the three symmetries which we will use to obtain flavor
predictions: the grand unified gauge symmetry and supersymmetry.

The successful prediction of $\sin^2\theta$ resulted from requiring a larger
symmetry than required by experiment.
It is well known that this  same enlargement of the gauge symmetry can also
yield predictions in the flavor sector.
Flavor observables at the weak scale, $F_a(M_W)$, can be given by relations of
the form
$$
F_a (M_W) = C_a \; F_G \;  \eta_a \eqno(2)
$$
where $C_a$ are again purely numerical group theory constants, while the
dynamical factors $\eta_a$ depend on several parameters, including
$\alpha_s$ and mass ratios such as $M_W/M_G$.
$F_G$ represents the set of independent flavor parameters of the GUT. Clearly a
predictive theory must have fewer such parameters than the 13 flavor parameters
of the standard model.
The first such prediction in GUTs was for $m_b/m_{\tau}$\cite{CEG}.
However, we now know that in this case $\eta_a$ depends on $m_t$ and
$\alpha_s$,
leading to uncertainties of 30\% and 10\% respectively.
Hence this successful prediction is much less significant than $\sin^2 \theta$,
especially as one successful prediction out of so many flavor parameters is not
convincing. The first successful prediction  of type (2) following from family
symmetries was $ \sin \theta_c = \sqrt{m_d/m_s}$ \cite{d/s}. While successful,
this is again a single relation at the 10\% accuracy level.

What level of significance can be expected in general from these type of flavor
predictions?
This is determined by the experimental uncertainties, both of the
predicted quantities and of the inputs used to
determine the free parameters of the theory. For example if the muon mass could
be predicted with only the electron mass needed as input, then the significance
would be extremely high. However, nobody is even close to being able to do
this.
In fact the best we are able to do is to use the six most accurately measured
flavor parameters as inputs: $m_e, m_\mu$ and $m_\tau$ are known at the 1 in
$10^3$ level or better, the Cabibbo angle to 1\%, and $m_c$ and $m_b$ to 5 -
10\%. In addition, to calculate the dynamical effects one needs to know the
strong gauge coupling, which is known at the 10\% level. Hence in this case the
level of significance of the predictions is dominated by how well the
predicted quantity is known, and this varies from around 15\% to 60\%. The
crucial lesson is that no single prediction
of this sort can possibly be very significant.
{\em The only hope that this approach will lead to significant successes is if
there are a large number of predictions.} Imposing grand unified, family and
super-symmetries still allows a vast number of possible theories. How are we to
decide which such theories to study? My answer is that we will simply study
those which offer the hope of obtaining the largest number of predictions
within a simple set of assumptions.
We are hoping that nature is kind to us and
that the flavor sector of the GUT depends on only a very few parameters.
While this `principle of maximal
predictivity' could be criticized as arbitrary, I would argue that
either nature is kind or the approach is not worth pursuing.

The power of combining family and GUT symmetries was realized by Georgi and
Jarlskog \cite{GJ} who wrote down a simple pattern for the Yukawa coupling
matrices at the GUT scale. This led to relations of the form of equation (2)
allowing successful predictions for all down type quark masses: $m_d, m_s$ and
$m_b$. Harvey, Ramond and Reiss \cite{HRR}
showed how to obtain this same pattern in the context of an
SO(10) GUT. They found that such a scheme violated CP, led to a prediction for
$m_t$ from $V_{cb}$ and allowed predictions to be made in the neutrino sector
as well. More recently Dimopoulos, Hall and Raby \cite{DHR} showed that this
Georgi-Jarlskog pattern, when used in a supersymmetric theory, was consistent
with everything we know about the flavor parameters. In addition to $m_d, m_s,
m_b$ and $m_t$, we found that the form of the KM matrix allows
$|V_{ub}/V_{cb}|$ and $J$ to be computed. We were astonished to find that
these two predictions and the top quark mass
prediction were all successful. While it is a
relatively simple matter to construct a theory which gets any one or two of
these relations it is very non-trivial to get all six simultaneously.
Despite the success of this framework I will not discuss the predictions
further. In the rest of this talk I describe a very different class of
SO(10) theories which have just 6 flavor parameters and are therefore
even more predictive \cite{ADHRS}.

The following assumptions are used to define this class of theories:
\begin{itemize}
\item The gauge group is SO(10). This is the smallest gauge group that allows
an entire family to be described by a single irreducible representation. Thus
the three families are written as $16_i$ with i=1,2,3 and $16_3$ being the
heaviest family. Perhaps the most elegant feature of SO(10) is the way in
which all
the measured gauge charges of the fermions can be simply understood in terms of
this 16 dimensional spinor.
\item The GUT is supersymmetric. Below the grand unification scale we take the
theory to be the minimal supersymmetric standard model, as this is the unique
minimal possibility for obtaining the successful $\sin^2 \theta$ prediction.
\item The two low energy Higgs doublets of this theory lie in a
single 10 dimensional representation of SO(10). This is the unique minimal
possibility.
\item The masses of the heavy generation ($m_t, m_b$ and $m_\tau$) come from a
single renormalizable operator
$$
A\  16_3\ 10\ 16_3 \eqno(3)
$$
where 10 is the multiplet containing the light doublets.
This elegant picture of the unification of the Yukawa couplings $\lambda_t$,
$\lambda_b$ and  $\lambda_\tau$ is reminiscent of the unification of the three
gauge couplings $g_1, g_2$ and $g_3$ and is due to Ananthanarayan, Lazarides
and Shafi \cite{ALS,ACKMPRW}.
\item All the masses of the quarks and leptons of the lightest two
generations, and the mixing angles of the KM matrix, are entirely due to
non-renormalizable operators which give masses suppressed compared to those
from (3) by powers of $M_G/M_P$ where $M_P$ is the Planck scale. Thus the mass
hierarchy between generations and the smallness of the KM angles is to be
understood in terms of powers of $M_G/M_P$.
We study those models with the fewest such
operators required for consistency with the known masses and mixings.
\item These non-renormalizable operators have the
form
$$
O_{ij} \equiv
16_i\, {45_1\over M_1}. . .  {45_k\over M_k}\, 10\, {45_{k+1}\over M_{k+1}} . .
 .
{45_\ell\over M_\ell}\, 16_j\eqno(4)
$$

The mass terms result when the various 45 dimensional adjoint representations
acquire vacuum expectation values (vevs) of order $M_G$.
\item Simple relations amongst the masses in the up, down and electron sectors
follow because each 45 vev lies in a definite direction in the SO(10) group
space: in the hypercharge, $B-L$, $T_{3R}$ or X direction, where X preserves an
SU(5) subgroup. To see this, recall that when a 45 vev acts on a fermion in a
16, it gives a numerical ``Clebsch'', which is the charge of the fermion
under the particular group generator
corresponding to the direction of this vev.
It is the SO(10) group theory ``Clebschs'' which allow an understanding of the
regularities of the fermion mass matrix \cite{DIM83}.
\item SO(10) may be broken to SU(5) by a vev of $45_X$ at a larger scale than
SU(5) is broken. This means that the objects appearing in the denominators in
(4) can be $<45_X>$ as well masses of order $M_P$.
\end{itemize}

At least two operators of the form (4) are needed in order to give all
quark and leptons a mass. Two such operators, together with (3), allow the $3
\times 3$ Yukawa matrices to have non-zero determinants. However the
coefficients of these three operators can all be made real by rotating the
phases of the three $16_i$ fields. Hence this case is excluded because the CP
violation in the KM matrix, $J$, vanishes.

The most predictive theories of this sort therefore have three operators of
type (4) in addition to the operator (3). Now only three of the four operator
coefficients can be made real, so that there are five independent GUT flavor
parameters. In addition, the quark and lepton masses depend on $\tan \beta$,
the
ratio of the two Higgs doublet vevs, so there are a total of six independent
flavor parameters. We choose to determine these from the six best measured
flavor parameters: $m_e, m_\mu, m_\tau, \theta_c, m_c$ and $m_b$. Hence the
theory predicts $\tan \beta$ and the seven standard model flavor parameters
$m_t, V_{cb}, m_s, m_d, m_u, V_{ub}$ and $J$.
We are currently performing a numerical search for all successful theories of
this type, and while the search has not yet been completed, we already know
that
there is a unique favored class of models. In this talk I will discuss only
this favored class, which is selected by the additional requirement that
\begin{itemize}
\item there must be a natural understanding
of why $m_c/m_t \approx V_{cb}^2 \ll m_s/m_b \approx m_\mu/m_\tau$ for
quantities renormalized at the GUT scale.
\end{itemize}

A lengthy but straightforward argument shows that the set of assumptions marked
above by bullets leads to Yukawa coupling matrices renormalized at the GUT
scale of the form:
$$
{\bf U} = \pmatrix{ 0&{1\over 27}C&0\cr
{1\over 27}C&0&x_uB\cr
0&x'_uB&A}
$$
$$
{\bf D} = \pmatrix{0&C&0\cr
C&Ee^{i\phi'}&x_dB\cr
0&x'_dB&A}$$
$$
{\bf E} = \pmatrix{ 0&C&0\cr
C&3Ee^{i\phi'}&x_eB\cr
0&x_e'B&A}
\eqno(5)
 $$
where A occurs in (3), and $B, C$ and $Ee^{i\phi'}$ are proportional to
the coefficients of the
three non-renormalizable operators of type (4), which must be chosen to
contribute to the 23, 12, and 22 entries of the matrices respectively. Notice
that while operator (3) yields $U_{33} = D_{33} = E_{33}$, a similar equality
is not found for the non-renormalizable contributions. For these the 45 vevs
introduce simple numerical Clebsch factors: $U_{22} : D_{22} : E_{22} = 0 : 1 :
3$ and $U_{12} : D_{12} : E_{12} = 1 : 27 : 27$. The 22 entry is infact the one
similarity of this scheme with the Georgi-Jarlskog pattern. While we have
proved that there is a unique successful Clebsch ratio for the 12, 21, and 22
entries, the 23 and 32 cases are quite different. Several Clebsch ratios are
possible and we have parameterized these discrete possibilities by $x_i$ and
$x_i'$ in eq. (5). Infact the low energy predictions depend on only two
combinations of these Clebsch parameters.

To demonstrate the power of these theories I will write down the analytic
formulas for the eight predictions. The predictions follow from relations of
the form of eq. (2). A technical problem is that the dynamical renormalization
group factors, $\eta_a$, depend not only on $\alpha_s$, but also on the third
generation Yukawa parameter $A$. Hence the determination of $A$, and of the
$\eta_a$, is a non-linear problem, which has no analytic solution. Of course A
and $\eta_a$ can be numerically computed with good accuracy. Hence I will give
the predictions in terms of the 6 input parameters, $A$ and $\eta_a$ and one
should simply remember that $A$ and $\eta_a$ are understood to be computed
numerically from the inputs.

Since the predictions are obtained from relations of the type of eq. (2) with
the GUT parameters $F_G$ determined from the inputs, the predictions take the
form
$$
\pmatrix{\hbox{predicted}\cr\hbox{quantity}} =
\pmatrix{\hbox{group theory}\cr \hbox{GUT Clebsch}}
\pmatrix{\hbox{input}\cr\hbox{parameter}}
\pmatrix{\hbox{dynamical}\cr \hbox{RG factor}}
\eqno(6)
$$
The eight predictions are as follows. The ratio of electroweak vevs $\tan
\beta$ is obtained from
$$
\cos\beta = { \sqrt{2}m_\tau\over v } \:{ \eta_1 \over A} \eqno(P1)
$$
where $v = 247$ GeV. The top quark mass parameter is
$$
m_t = \tan\beta \ m_\tau\: \eta_2 \eqno(P2)
$$
with $\beta$ determined from eq. (P1). These two predictions follow from just
operator (3) for the heaviest generation and in these cases the GUT Clebsch
factor is unity. The mixing between the two heaviest
generations is given by
$$
V_{cb} = \chi \sqrt{ {m_c\over m_t}} \: \eta_3 \eqno(P3)
$$
Thus the GUT relation $V_{cb} = \sqrt{ {m_c\over m_t}}$ \cite{HRR,DHR,Vcb} is
modified by a Clebsch factor $\chi$, which we discuss below.
Whenever $m_t$ appears on the right-hand side of a prediction, it is understood
that the value given by (P2) is to be used.
The strange mass is given by
$$
m_s= {1\over 3}(1+\delta)  {m_b\over m_\tau} m_\mu \: \eta_4\eqno(P4)
$$
which is one of
the Georgi-Jarlskog relations, except for a small correction $1+\delta$
which we discuss below.
{}From the determinant of {\bf D} and {\bf E} one finds
$$
{m_s\over m_d} = {1\over 9} (1+2\delta)  {m_\mu\over m_e}\eqno(P5)
$$
which is a small modification of the second Georgi-Jarlskog relation. In this
prediction the renormalization factors cancel.
The prediction for $m_u/m_d$ follows from the determinants of the Yukawa
matrices,
with $m_s$ substituted from (P4):
$$
{m_u\over m_d} = {1\over 3^7}(1+\delta)
 {m_\mu\over m_\tau} { m^2_t\over m_cm_b}
 \eta_6 \eqno(P6)
$$

The last two predictions are for parameters of the KM matrix. Diagonalization
of {\bf U} and {\bf D} yield a KM matrix of the form
$$
V =
\pmatrix{ c_1c_2-s_1s_2e^{-i\phi}&s_1+c_1s_2e^{-i\phi}&s_2s_3\cr
-c_1s_2-s_1e^{-i\phi}& c_1c_2c_2 e^{-i\phi}-s_1s_2& c_2s_3\cr
s_1s_3&-c_1s_3& c_3e^{i\phi} }
\eqno(7)
$$
where $s_1$, $s_2$ and $\phi$ are renormalization group invariants, and
$s_3 = V_{cb}$ has a simple scaling behaviour. The CP violating phase $\phi$ is
derived from, but not identical to, the phase $\phi'$ of eq. (5). The angles
$s_1$ and $s_2$ are given by
$$
s_1 = \sqrt{{m_d \over m_s}} \eqno(8)
$$
$$
s_2 = \sqrt{{m_u \over m_c}} \eqno(9)
$$
and $\phi$ is determined from the Cabibbo angle via
$$
\sin \theta_c = |V_{us}| = | s_1+ c_1 s_2 e^{-i\phi}|\eqno(10)
$$
The two quantities of {\bf V} which are predicted are
$$
|{V_{ub}\over V_{cb}}| = s_2 \eqno(11)
$$
and the amount of CP violation
$$
J = s_1 s_2 s_3^2 s_\phi  \eqno(12)
$$
It is very interesting to note that equations (7) - (12) also hold in the
Georgi-Jarlskog scheme \cite{DHR}. Indeed it has recently been shown that the
successful predictions (11) and (12) follow from very simple assumptions about
the form of {\bf U} and {\bf D} \cite{HR}. However, the class of theories
under study here is much more predictive and makes specific predictions for
$ \sqrt{{m_d \over m_s}}$ and $ \sqrt{{m_u \over m_c}}$ which can then be
substituted in eq. (8) and (9). The prediction for $ \sqrt{{m_d \over m_s}}$ is
obtained from (P5) while the prediction for $ \sqrt{{m_u \over m_c}}$ gives
$$
|{V_{ub}\over V_{cb}}| = s_2
 = {1\over 27} {m_e^{1/2}m_\mu^{1/2}\over m_\tau}
{m_t\over m_c} \: \eta_7 \eqno(P7)
$$
The final prediction is for the amount of CP violation in the KM matrix
obtained by using the above expressions for $s_1, s_2$ and $s_3= V_{cb}$
in eq. (12)
$$
J = {\chi^2\over 9} (1-\delta) {m_e\over m_\tau} s_\phi \: \eta_8 \eqno(P8)
$$
where $\phi$ is obtained from (10).

The class of models under discussion does not have a unique operator
contributing to the 23 and 32 entries of the Yukawa matrices. This is reflected
in (5) by the appearance of the Clebschs $x_i$ and $x_i'$ which
can assume a set of
discrete values. Nevertheless all models of this class lead to the above 8
predictions and the only dependence on these Clebschs is through the two
parameters
$$
\chi = {x_u - x_d \over \sqrt{x_u x_u'}}\eqno(13)
$$
which only enters the relation for $V_{cb}$ and
$$
\delta = {x_ex_e'-3x_dx_d'\over x_ux_u'}
          {m_\tau m_c\over m_\mu m_t} \; \eta_\delta \eqno(14)
$$
The prediction (P3) impies that the only theoretically allowed
values of $\chi$ which are experimentally acceptable are: $\chi = 2/3,
5/6$ and $8/9$. The case $\chi=1$ \cite{HRR,DHR,Vcb} is disfavored in the
present theories which contain operator (3) because the resulting values
for $V_{cb}$ are uncomfortably large.
For all models of interest $\delta \ll 1$ and hence the $\delta$ dependence
of $m_s$ in (P4) and of $J$ in (P8) is much less than the experimental
uncertainties on these quantities, and can be dropped. The more interesting
effects of $\delta$ are to be found in (P5) and in (P6), where they give small
modifications to the ratios $m_u/m_d$ and $m_s/m_d$.

In summary our eight predictions for $\beta, m_t, V_{cb}, m_s, m_s/m_d,
m_u/m_d, V_{ub}/V_{cb}$ and $J$ are given in (P1) - (P8).
They all agree with present experimental values, and the predictions are
sufficiently accurate that future experiments will provide critical tests of
these theories. The most important advances which can test our scheme via
predictions (P1) - (P8) are
\begin{itemize}
\item A measurement of $m_t$.
\item A high statistics study of semi-leptonic B meson decay to measure
$V_{cb}$. In addition, better theoretical understanding of this matrix element
is required, which looks likely in view of recent developments in heavy quark
effective field theory.
\item A measurement of the CP violating decays of neutral B mesons, which will
test our predictions for the KM matrix.
\item A better theoretical understanding of the values for $m_u/m_d$ and
$m_s/m_d$ implied by experiment.
\end{itemize}

There are two obvious objections to the above scheme:

1) While there are only six independent continuous flavor parameters, there are
   millions of operators of the form of (4), and therefore there
   are extra discrete variables: the Clebschs. If Clebschs can be found to fit
   any values of the standard model parameters, then there is no significance
to
   our results.

Infact we find the set of possible Clebschs to be very coarse-grained.
As Clebschs are varied from one set to the next set, the value of a
predicted quantity is found to jump by amounts typically much larger than its
experimental error bar, hence successful predictions are significant. A case
where this is not true is the prediction (P3) for $V_{cb}$. In this case the
experimental error bar is of order the interval generated by successive
possible values of the Clebsch $\chi$. A modest decrease in the experimental
error bar will simply serve to choose one of the three presently allowed
values of $\chi$.

2) Our scheme is based on a large number (9) of assumptions, suggesting that it
   is unlikely to be the one chosen by nature.

My response to this is mixed. It may well be that the ``zeroth order''
assumption is wrong and that this whole approach to fermion masses is
incorrect. However, I have argued that this is the only known approach which
yields predictions of any significance which can be compared to experiment.
Given that this approach is worth pursuing, I would argue that the set of 9
assumptions which we have made is {\em the simplest} that leads to models of
such {\em high predictivity}. There are undoubtedly more complicated sets of
assumptions, and obviously there are less predictive theories, but
{\em without major additions to the basic tools it is unlikely that there is a
simpler, more predictive model}. The success of the predictions gives me
optimism that nature may have chosen the very simplest direction.

SO(10) grand unified theories offer the hope that neutrino masses can be
predicted once a sufficiently simple flavor sector has been written down. This
is because both left and right-handed neutrinos lie in the same 16 dimensional
spinor representation as the quarks and charged leptons. Actually, it is the
neutrino mass {\em ratios} and the leptonic mixing angles which can be
accurately determined. The overall neutrino mass scale involves knowing the
scale of lepton number violation responsible for the right-handed neutrino
Majorana masses, and this cannot be determined from charged lepton or quark
masses. The implementation of the Georgi-Jarlskog ansatz in SO(10) can yield
specific forms for the neutrino mass matrices \cite{HRR}. Two very specific
forms predict all the mass ratios and mixing angles as shown in the
Table
\cite{DHRnu}:\\

\begin{center}
\begin{tabular}{|c|c|c|}
\hline
&I&II\cr
\hline
$\theta_{e\mu}$ & $(6.5\pm .3) 10^{-2}$&$.15\pm .04$\cr
$\theta_{\mu\tau}$&$.081\pm .008$&$-.027\pm .003$\cr
$\theta_{e\tau}$&$(5.7\pm .6) 10^{-4}$&$(1.9\pm 0.2)10^{-4}$\cr
$m_{\nu_\tau}/m_{\nu_\mu}$&$208 \pm 42$&$1870\pm 370$\cr
$m_{\nu_\mu}/m_{\nu_e}$&$(3.1\pm 1.0) 10^3$&$38 \pm 12$\cr
$m_{\nu_{\tau max}}$& 2.5 eV&710 eV\cr
\hline
\end{tabular}
\end{center}

\centerline{\bf Table 1}
\vskip .5in

While extra assumptions, beyond those of Georgi and Jarlskog, are required to
obtain these numbers, the very fact that such precise predictions can be made
is an important result in itself.
The best hope for testing these neutrino masses is offered by
searches for $\nu_\mu \nu_\tau$ oscillations by the CHORUS and NOMAD
experiments at CERN and by P803 at Fermilab \cite{nuos}.
The SO(10) scheme which leads to predictions (P1) - (P8) cannot be directly
used to predict quantities of the neutrino sector. Substantial modifications
are required, and these are presently being studied.

Extraordinary effort is involved in measuring the 18 parameters of the standard
model. Why bother? Two answers are frequently given:
\begin{itemize}
\item because they are there and they are fundamental.
\item By measuring them more accurately, via a variety of methods, one could
uncover inconsistencies in the standard model which would indicate new physics.
\end{itemize}
While both of these arguments have considerable merit, a third reason is also
important:
\begin{itemize}
\item the accurate determination of the 18 parameters of the standard model may
lead us to a deeper understanding of particle physics: we may be led to a
predictive theory behind the standard model in the same way that atomic spectra
were crucial in pointing the way to the Bohr model and to quantum mechanics.
\end{itemize}

I have argued that we have all the symmetry tools we need
to construct predictive theories of fermion masses.
Should this direction be correct, does this mean we
have no need of a revolution in the underlying theory?
Quite the reverse: parameter reduction gets us far along the road, but it
cannot be the whole story. Eventually we do need a new framework to address
such questions as why the symmetries are what they are and why the free
parameters (there are always some) take the observed values. However, it may be
possible to go very far down the road guided by experiment, and unexpected
features of the underlying theory may only then become apparent.


\begin{thebibliography}{99}

\bibitem{WGG} S. Weinberg {\it Phys. Rev.} {\bf D 5} 1962 (1972);
H. Georgi and S. Glashow {\it Phys. Rev.} {\bf D 7} 2457 (1973).
\bibitem{etc} S. Dimopoulos and L. Susskind {\it Nucl. Phys.} {\bf B155} 237
(1979); E. Eichten and K. Lane, {\it Phys. Lett.} {\bf 90B} 125 (1980).
\bibitem{string}P. Candelas, {\it Nucl. Phys.} {\bf B298} 458 (1988);
B.R. Greene, K. H. Kirklin, P.J. Miron and G. G. Ross, {\it Phys. Lett.}
{\bf 192B} 111 (1987); {\it Nucl. Phys.} {\bf B292} 606 (1987).
\bibitem{tcond} Y. Nambu, EFI 88-39, EFI 88-62 (1988); V. Miranski, M.
Tanabashi and K. Yamawaki, {\it Mod. Phys. Lett.} {\bf A 4} 1043 (1989); W. J.
Marciano, {\it Phys. Rev. Lett.} {\bf 62} 2793 (1989); W. A. Bardeen, C. T.
Hill and M. Lindner {\it Phys. Rev.} {\bf D 41} 1647 (1990).
\bibitem{GG} H. Georgi and S.L. Glashow,
{\it Phys. Rev. Lett.} {\bf 32} 438 (1974).
\bibitem{GQW} H. Georgi, H. Quinn and S. Weinberg,
{\it Phys. Rev. Lett.} {\bf 33} 451 (1974).
\bibitem{susys2} S. Dimopoulos, S. Raby and F. Wilczek, {\it Phys. Rev .}
{\bf D24} 1681 (1981); S. Dimopoulos and H. Georgi, {\it Nucl. Phys.}
{\bf B193},150 (1981); L. Ibanez and G. G. Ross, {\it Phys. Lett.} {\bf 105B}
439 (1981); M. B. Einhorn and D. R. T. Jones {\it Nucl. Phys.} {\bf B196} 475
(1982); W. J. Marciano and G. Senjanovic, {\it Phys. Rev.} {\bf D 25} 3092
(1982).
\bibitem{gutth} S. Weinberg, {\it Phys. Lett} {\bf 91B} 51 (1980); L. J. Hall,
{\it Nucl. Phys.} {\bf B178} 75 (1981);
R. Barbieri and L.J. Hall, {\it Phys. Rev. Lett.} {\bf 68} 752 (1992).
\bibitem{planckth} C. T. Hill, {\it Phys. Lett} {\bf 135B} 47 (1984); L. J.
Hall and U. Sarid, LBL-32905 (1992).
\bibitem{weakth}G. G. Ross and R. G. Roberts {\it Nucl. Phys.} {\bf B377} 571
(1992);
J. Hisano, H. Murayama and T. Yanagida, {\it Phys. Rev. Lett.}
{\bf 69} 1014 (1992);
R. Arnowitt and P. Nath, {\it Phys. Rev. Lett.} {\bf 69} 725 (1992);
S. Kelley, J. L. Lopez, D. V. Nanopoulos, H. Pois and K. Yuan, CERN TH6498/92
(1992);
P. Langacker and N. Polonsky, UPR-0513T (1992).
\bibitem{CEG} M. Chanowitz, J. Ellis and M.K. Gaillard {\it Nucl. Phys.}
{\bf B135} 66 (1978).
\bibitem{d/s} S. Weinberg, A Festschrift for I.I. Rabi, Transactions of the New
York Academy of Sciences (1977), F. Wilczek and A. Zee; {\it Phys. Lett.} {\bf
B70} 418 (1977); H. Fritzsch, {\it Phys. Lett.} {\bf B70} 436 (1977).
\bibitem{GJ} H. Georgi and C. Jarlskog, {\it Phys. Lett. } {\bf 86B} 297
(1979).
\bibitem{HRR} J. Harvey, D. Reiss and P. Ramond, {\it Phys. Lett.} {\bf 92B}
309
(1980); {\it Nucl. Phys.} {\bf B199} 223 (1982).
\bibitem{DHR} S. Dimopoulos, L.J. Hall and S. Raby, {\it Phys. Rev. Lett.}
{\bf 68} 1984 (1992), {\it Phys. Rev.} {\bf D45} 4192 (1992), {\it Phys. Rev.}
{\bf D 46} R4793 (1992).
\bibitem{ADHRS} G. Anderson, S. Dimopoulos, L. J. Hall, S. Raby and G.
Starkman, in preparation (1993),
\bibitem{ALS} B. Ananthanarayan, G. Lazarides and Q. Shafi, Phys. Rev. {\bf
D44}
1613 (1991).
\bibitem{ACKMPRW} H. Arason, D. Casta\~no, B. Keszthelyi, S. Mikaelian,
E. J. Piard, P. Ramond and B. Wright, Phys. Rev. Lett. {\bf 67}
2933 (1991).
\bibitem{DIM83} S. Dimopoulos, Phys. Lett. {\bf 129 B} 417 (1983).
\bibitem{Vcb} P. Ramond, UFIFT-92-4 (1992); H. Arason, D.  Casta\~no, E. J.
Piard
and P. Ramond, {\it Phys. Rev.} {\bf D 47} 232 (1993).
\bibitem{HR} L. J. Hall and A. Ra\v{s}in, LBL 33668 (1993).
\bibitem{DHRnu} S. Dimopoulos, L.J. Hall and S. Raby, LBL 32484 (1992).
\bibitem{nuos}CHORUS collaboration, N. Armenise et al., CERN SPSC/90-42 (1990);
NOMAD collaboration, P. Astier et al., CERN SPSC/91-21 (1991); Fermilab P803
proposal, K. Kodama et al., (1990).
%

\end{thebibliography}
\end{document}